\documentstyle[11pt]{article}	
\advance\textwidth	.4in
\advance\oddsidemargin	-.2in
\advance\textheight	.5in
\advance\topmargin	-.25in
\def\pl{\partial}
\def\scst{\scriptstyle}
\def\diag{\mathop{\rm diag}}
\def\tr{\mathop{\rm tr}}	
\def\t{\mathopen{{}^t}}		

\title{Calogero-Moser hierarchy and KP hierarchy}
\author{Takahiro Shiota\\Department of Mathematics, Kyoto University\\
Kyoto 606--01, JAPAN}
\date{}

\begin{document}
\maketitle

\section{Introduction}

In \cite{amm}, Airault, McKean and Moser observed that the motion of
poles of a rational solution to the K-dV or Boussinesq equation obeys
the Calogero-Moser dynamical system \cite{calogero,calogero2,calogero3}
with an extra condition on the configuration of poles.
In \cite{krchvr8}, Krichever observed that the motion of poles of a
solution to the KP equation which is rational in $t_1$ obeys the
Calogero-Moser dynamical system.
In this note we shall generalize those results to the KP {\em hierarchy}.
Noting that a pole of a KP solution comes from a zero of the corresponding
$\tau$-function, the statement becomes thus

\proclaim Theorem.
	$(1)$ Let $\tau(t_1,t_2,t_3)$ be a monic polynomial in $t_1$ whose
coefficients depend on $t_2$ and $t_3$.  If $\tau$ satisfies the KP equation
\begin{equation}
(D_1^4 + 3D_2^2 - 4D_1D_3)\tau\cdot\tau = 0,
\label{KPeqn}
\end{equation}
then $\tau$ can be extended uniquely to a $\tau$-function of the KP hierarchy
which is also a monic polynomial in $t_1$.\hfill\break
	$(2)$ Let $\tau=\tau(t)$, where $t=(t_1,t_2,\dots)$, be a monic
polynomial in $t_1$ whose coefficients depend on $t':=(t_2,t_3,\dots)$.
Then $\tau$ is a $\tau$-function of the KP hierarchy if and only if the
motion of zeros of $\tau$ is governed by the hierarchy of Calogero-Moser
dynamical systems.  Namely, writing
\begin{equation}
	\tau(t)=\prod_{i=1}^N(t_1-x_i(t'))
\label{x_i}
\end{equation}
locally in $t'$, $\tau$ is a $\tau$-function if and only if
\begin{equation}
{\pl\over\pl t_n}\biggl(\begin{array}{c}x_i\\\xi_i\end{array}\biggr) =
(-1)^n	\biggl(\begin{array}{r}\pl H_n/\pl\xi_i\\-\pl H_n/\pl x_i
	\end{array}\biggr),
\quad n=2,3,\dots,
\label{cm}
\end{equation}
where $\xi_i=(1/2)\pl x_i/\pl t_2$, $H_n=\tr Y^n$, and $Y$ is the
Moser matrix, i.e., $Y=Y(t')$ is an $N\times N$ matrix such that
\begin{equation}
Y_{ij}={1\over x_i-x_j}\quad(i\ne j),\qquad Y_{ii}=\xi_i.
\label Y
\end{equation}

Note that $x_i(t')\ne x_j(t')$ for $i\ne j$ and almost every $t'$ since the
zeros of $\tau$ as a function of $t_1$ is generically simple
(see, e.g., \cite[Prop.~8.6]{sw}).

This class of KP solutions (let us call them {\it quasi-rational\/}) might
not have drawn so much attention as the rational solutions, the ones
whose $\tau$-functions are polynomials in all time variables.  Rational
solutions have been studied in detail as they play the central role in the
Kyoto school's theory of soliton equations.
Recent developments in the study of the Calogero-Moser hierarchy may lead
to a better understanding of quasi-rational solutions.

Rational solutions are clearly\footnote{
	For a $\tau$-function which is a polynomial in $t_1$, it is easy to
	observe that the coefficient of the top degree term in $t_1$ does not
	vanish.  In particular, the coefficient is a constant if $\tau$ is a
	polynomial in all variables.
}
quasi-rational, but the converse is not necessarily true.  The difference
between the rational and quasi-rational solutions, and other properties of
quasi-rational solutions, can be seen easily from the following very explicit
formula:

\proclaim Corollary 1. Let $\tau$ be the $\tau$-function of a quasi-rational
solution of the KP hierarchy (i.e., $\tau$ is a monic polynomial in $t_1$
with variable coefficients).  Using the notation as in part $(2)$ of the
theorem, and assuming for simplicity that $x_i(0)\ne x_j(0)$ for $i\ne j$,
we have
\begin{equation}
\tau(t)=\det\biggl(-X_0+\sum_{n=1}^\infty (-1)^{n-1}nt_n(Y_0)^{n-1}\biggr),
\label{explicit}
\end{equation}
where $X_0=\diag(x_1(0),\dots,x_N(0))$, and $Y_0$ is the Moser matrix at
$t'=0$, i.e.,
$$
(Y_0)_{ij}={1\over x_i(0)-x_j(0)}\quad(i\ne j),\qquad Y_{ii}=\xi_i(0).
$$

This follows immediately from part (2) of the theorem and a well-known
formula on the Calogero-Moser hierarchy.  It implies, for example, that a
quasi-rational solution is a rational solution if and only if $Y_0$ is
nilpotent, and that the $\tau$-function of a quasi-rational solution is
actually a polynomial in any first finitely many time variables.  Hence we
have

\proclaim Corollary 2.  A solution to the KP hierarchy is quasi-rational if
and only if it is a finite dimensional solution \cite{krchvr4,mfd:agc,sw,sht}
whose spectral curve $C$ is a rational curve (i.e., its normalization is the
Riemann sphere) with only cusp-like singularities (possibly including higher
cusps).

Indeed, from (\ref{explicit}) the orbit of a quasi-rational solution is a
finite dimensional affine space.  Since the orbit is isomorphic to the
(generalized) Jacobian of $C$, $C$ is a rational curve with only cusp-like
singularities.
Conversely, the $\tau$-function belonging to a rational spectral curve with
cusp-like singularities gives a polynomial on the Jacobian as a regular
function on the affine space, proving the corollary.

A closer look reveals that the singularities of $C$ are at the zeros of
$\det(z + Y_0)$, i.e., the negative of the eigenvalues of $Y_0$, where $z$
is the global coordinate on the normalization of $C$ given by the analytic
extension of the formal spectral parameter in the Baker-Akhiezer function
(\ref{BA}).

There should be no essential difficulties in generalizing our theorem to the
trigonometric or elliptic Calogero-Moser hyerarchy, except that in the
elliptic case the arguments relying on the vanishing of logarithmic
derivatives of $\tau$ at $t_1=\infty$ no longer work.
The trigonometric or elliptic analogue of corollary 2, e.g., to study the
geometry of spectral curves which give KP solutions periodic in $t_1$,
may be less trivial.  A. Treibich and J.-L. Verdier have some results in this
direction \cite{tv}.  After completion of this work, the author found out
that Alex Kasman \cite{ak} has similar work.

\section{Proof of theorem}

Throughout, we use the standard notation for the KP hierarchy.  We shall
denote $\pl_x=\pl/\pl x$ and $\pl_i=\pl/\pl t_i$.  By abuse of notation,
we shall ``identify" $t_1$ and $x$, and simply denote $t_1+x$ by $t_1$.
Our proof is similar to the extension procedure of \cite[Sect.~3]{sht}.
Starting from Krichever's polynomial solution of the KP equation
\cite{krchvr8}, we shall construct the Baker-Akhiezer function in such a
way that the resulting extension of $\tau$ remains to be a polynomial in
$t_1$.  Then we shall observe
that the time evolution of zeros is governed by the Calogero-Moser hierarchy.

First assume $\tau=\tau(t_1,t_2,t_3)$ is of the form (\ref{x_i}), except that
$t'$ is replaced by $(t_2,t_3)$, and
satisfies the KP equation.  Krichever \cite{krchvr8} observed that $x_i$
and $\xi_i=(1/2)\pl_2x_i$ satisfy the equations (\ref{cm}) for $n=2$,~3.
In particular, we have
\begin{equation}
\pl_2\xi_i=-{\pl\over\pl x_i}\tr Y^2
=-4\sum_{j(\ne i)}{1\over(x_i-x_j)^3}.
\label{pet}
\end{equation}

To extend $\tau$ to a $\tau$-function of the whole KP hierarchy, we
construct the Baker-Akhiezer (wave) function
\begin{equation}
w = \biggl(\sum_{m=0}^\infty w_mz^{-m}\biggr)e^{\sum t_iz^i},
\label{BA}
\end{equation}
where $w_m=w_m(t)$ and $w_0\equiv1$.
Since $\tau$ is of the form (\ref{x_i}), we have
$$
u_2:=\pl_1^2\log\tau
=\sum_{i=1}^N{-1\over(t_1-x_i)^2},
$$
so that the equation $\pl_2 w = B_2 w$, $B_2:=\pl_x^2+2u_2$, becomes
\begin{equation}
\pl_1 w_{m+1}
={1\over2}(\pl_2-\pl_x^2)w_m + \sum_{i=1}^N{1\over(t_1-x_i)^2}w_m,
\quad m=0,1,\dots.
\label{rr}
\end{equation}
Since after the desired extension we should have
$$
w_m={p_m(-\tilde\pl)\tau\over\tau},\quad m\ge0
$$
(see \cite{djkm}), $w_m$, $m>0$, should have at most simple poles along the
zero locus of $\tau$, and vanishes at infinity:
\begin{eqnarray}
w_{m+1}\bigr|_{t_1=\infty}=0,\quad m=0,1,\dots.
\label{bc}
\end{eqnarray}
Given $w_m$, (\ref{rr}) and (\ref{bc}) determine $w_{m+1}$ uniquely.
We shall show that for $m\ge0$, $w_{m+1}$ has
at most simple poles, i.e., it has the form
\begin{equation}
w_{m+1}=\sum_{i=1}^N{c_{m+1,i}\over t_1-x_i},\quad m\ge0.
\label{poles}
\end{equation}
For $m=0$, the right hand side of (\ref{rr}) becomes $\sum_i1/(t_1-x_i)^2$
since $w_0\equiv1$.  Hence
\begin{equation}
c_{1,i}=-1.
\label{initp}
\end{equation}
For $m>0$, the coefficient of $(t_1-x_i)^{-2}$ in (\ref{rr}) gives
the recurrence relation
\begin{eqnarray}
-c_{m+1,i}&=&{1\over2}(\pl_2x_i)c_{m,i}+\sum_{j(\ne i)}{c_{m,j}\over x_i-x_j}
\nonumber\\
&=&\xi_ic_{m,i}+\sum_{j(\ne i)}{c_{m,j}\over x_i-x_j},
\label{rrp}
\end{eqnarray}
and the coefficient of $(t_1-x_i)^{-1}$ in (\ref{rr}) gives the constraint
\begin{equation}
0 = {1\over2}\pl_2c_{m,i}+\sum_{j(\ne i)}{c_{m,i}-c_{m,j}\over(x_i-x_j)^2}.
\label{single}
\end{equation}
Let ${\bf c}_m=\t(c_{m,1},\dots,c_{m,N})$, and
${\vec{\bf 1}}=\t(1,\dots,1)$.
 From (\ref{initp}) and (\ref{rrp}) we have
\begin{equation}
{\bf c}_m=(-1)^mY^{m-1}{\vec{\bf 1}},
\label c
\end{equation}
where $Y$ is the Moser matrix as in (\ref Y).
The condition (\ref{single}) is automatically satisfied.
This follows from the local regularity of a $\tau$-function
\cite[p.~366, lemma~8]{sht} (since otherwise $w_{m+1}$, and hence the
extended $\tau$, would be multivalued),
but in the present case we have a more direct proof:
 From (\ref Y), (\ref{pet}) and the definition of $\xi_i$,
we have the well-known relation
$$
\pl_2Y=2[Z,Y],\quad\hbox{where}\quad
Z_{ij}=\left\{
	\begin{array}{ll}
	1/(x_i-x_j)^2 & (i\ne j)\\[5pt]
	-\sum_{k(\ne i)}1/(x_i-x_k)^2 & (i=j)
	\end{array}
\right.
$$
Applying this to (\ref c), and noting $Z{\vec{\bf1}}=0$, we have
\begin{equation}
{1\over2}\pl_2{\bf c}_m
={(-1)^m\over2}\pl_2(Y^{m-1}){\vec{\bf 1}}
=(-1)^m[Z,Y^{m-1}]{\vec{\bf 1}}
=(-1)^mZY^{m-1}{\vec{\bf 1}}
=Z{\bf c}_m,
\label{d2c}
\end{equation}
which is equivalent to (\ref{single}).

Now we need a little formal nonsense.  We extend $w$ formally to the
whole KP hierarchy by solving the equations
\begin{equation}
\pl_nw=(W\pl_x^n W^{-1})_+w
\label{wevol}
\end{equation}
for $n=3$,~4,\,\dots, where $W=\sum_{m=0}^\infty w_m\pl_x^{-m}$.
Since those equations are compatible with one another, $w$ extends uniquely
to the Baker-Akhiezer function of the whole KP hierarchy.  We shall denote
the extended $w$, $w_m$ and $\tau$ by $\bar w$, $\bar w_m$ and $\bar\tau$,
respectively.  From (\ref{poles}) and (\ref{wevol}), $\bar w_m$ are also
single valued, and vanish at $t_1=\infty$.  Hence, since
$-\pl_1\log\bar\tau=\bar w_1=O(z^{-1})$, $\bar\tau$ has a polynomial growth,
so that it has a pole at $t_1=\infty$.  Also, each zero of $\tau$ extends to
a zero of $\bar\tau$ (see \cite[lemma~8]{sht}).
Hence $\bar\tau$ is a polynomial in $t_1$.  Since
$$
-(1/m)\pl_m\log\bar\tau + {P(\pl_1,\dots,\pl_{m-1})\bar\tau\over\bar\tau}
=\bar w_m=O(z^{-1})
$$
for some constant coefficient polynomial $P$, we observe by induction in $m$
that the top degree coefficient of $\bar\tau$ is constant in $t$, so that
$\bar\tau$ is a monic polynomial.
This proves part (1) of the theorem, and, combined with \cite{krchvr8},
proves that there is no constraints that $x_i(0)$ and $\xi_i(0)$ must
satisfy.  To complete the proof of part (2), we have only to determine
the time evolution of $x_i$.

Let $w^*=\bigl(\sum w_m^*z^{-m}\bigr)e^{-\sum t_iz^i}$, $w_0^*=1$, be the
adjoint Baker-Akhiezer function of $w$, i.e., $\pl_2w^* = -B_2^*w^*=-B_2w^*$,
whose coefficients $w_m^*$, $m>0$, vanish at $t_1=\infty$.  Setting
$$
w_{m+1}^*=\sum_{i=1}^N{c_{m+1,i}^*\over t_1-x_i},\quad m\ge0,
$$
the same calculations as above show
\begin{equation}
{\bf c}_m^*=(-1)^{m-1}\cdot\t{\vec{\bf 1}}Y^{m-1},\quad\hbox{and}\quad
{1\over2}\pl_2{\bf c}_m^*=(-1)^m\cdot\t{\vec{\bf 1}}Y^{m-1}Z,
\label{cstar}
\end{equation}
where ${\bf c}_m^*=(c_{m,1}^*,\dots,c_{m,N}^*)$.
Now we compute the coefficients of $\pl_x^{-1}$ in the equation
\begin{equation}
\pl_nW=-(W\pl_x^nW^{-1})_-W,
\label{Wevol}
\end{equation}
which is equivalent to (\ref{wevol}).  The left hand side gives
\begin{equation}
\pl_nW	=-\Bigl(\pl_1\pl_n\log\prod(t_1-x_i(t'))\Bigr)\pl_x^{-1}+O(\pl_x^{-2})
	=-\sum{\pl_nx_i\over(t_1-x_i)^2}\pl_x^{-1}+O(\pl_x^{-2}).
\label{lhsWevol}
\end{equation}
Comparing the formulas
$w^*=\bigl(\sum w_m^*(-\pl_x)^{-m}\bigr)e^{-\sum t_iz^i}$ and
$w^*=\t W^{-1}e^{-\sum t_iz^i}$ (see \cite{djkm}), we have
$$
W^{-1}=\sum\pl_x^{-m}\circ w_m^*,
$$
and hence the right hand side of (\ref{Wevol}) becomes
\begin{eqnarray*}
-(W\pl_x^nW^{-1})_-W
&=&-\biggl(\sum_{m+m'>n}w_m\pl_x^{n-m-m'}\circ w_{m'}^*\biggr)
(1+O(\pl_x^{-2}))
\\
&=&-\sum_{m+m'=n+1}w_mw_{m'}^*\pl^{-1}+O(\pl_x^{-2})
\\
&=&-\sum_{\scst m+m'=n+1\atop\scst m,m'>0}\biggl(
\sum_{i=1}^N{c_{m,i}c_{m',i}^*\over(t_1-x_i)^2} +\hbox{simple poles}
\biggr)\pl^{-1}+O(\pl_x^{-2})
\end{eqnarray*}
Comparing this with (\ref{lhsWevol}), and using (\ref c), (\ref{cstar}) and
$\pl Y/\pl\xi_i=E_{ii}$, where $E_{ij}$ denotes the $(i,j)$th matrix element,
i.e., $(E_{ij})_{kl}=\delta_{ik}\delta_{jl}$, we have
\begin{eqnarray}
\pl_nx_i
&=&\sum_{\scst m+m'=n+1\atop\scst m,m'>0}c_{m,i}c_{m',i}^*
 = \sum_{\scst m+m'=n+1\atop\scst m,m'>0}{\bf c}_{m'}^*E_{ii}{\bf c}_m
\nonumber\\
&=&(-1)^n\sum_{\scst m+m'=n+1\atop\scst m,m'>0}
(1,\dots,1)Y^{m'-1}E_{ii}Y^{m-1}
\left(\begin{array}c1\\\vdots\\1\end{array}\right)
\nonumber\\
&=&(-1)^n(1,\dots,1){\pl Y^n\over\pl\xi_i}
\left(\begin{array}c1\\\vdots\\1\end{array}\right)
\nonumber\\
&=&(-1)^n{\pl\over\pl\xi_i}\tr\left(
Y^n\left(\begin{array}c1,\dots,1\\\dots\dots\\1,\dots,1\end{array}\right)
\right)
\nonumber\\
&=&(-1)^n{\pl\over\pl\xi_i}\tr\bigl(Y^n([X,Y]+I)\bigr),
\label{cm1}
\end{eqnarray}
where $X=\diag(x_1,\dots,x_N)$, and $I$ is the identity matrix.  Since
$\tr(Y^n[X,Y])=\tr[X,Y^n]=0$, (\ref{cm1}) gives the first equation in
(\ref{cm}).

Finally, differentiating the first line of (\ref{cm1}), and applying
(\ref c), (\ref{d2c}), (\ref{cstar}) and $\pl Y/\pl x_i=-[E_{ii},Z]$, we have
\begin{eqnarray*}
\pl_n\xi_i={1\over2}\pl_2\pl_nx_i
&=&\sum_{\scst m+m'=n+1\atop\scst m,m'>0}({\bf c}_{m'}^*E_{ii}Z{\bf c}_m
-{\bf c}_{m'}^*ZE_{ii}{\bf c}_m)
\\
&=&(-1)^n\sum_{\scst m+m'=n+1\atop\scst m,m'>0}
(1,\dots,1)Y^{m'-1}[E_{ii},Z]Y^{m-1}
\left(\begin{array}c1\\\vdots\\1\end{array}\right)
\\
&=&-(-1)^n(1,\dots,1){\pl Y^n\over\pl x_i}
\left(\begin{array}c1\\\vdots\\1\end{array}\right)
\\
&=&-(-1)^n{\pl\over\pl x_i}\tr\left(
Y^n\left(\begin{array}c1,\dots,1\\\dots\dots\\1,\dots,1\end{array}\right)
\right)
\\
&=&-(-1)^n{\pl\over\pl x_i}\tr Y^n
	\qquad\hbox{(as in the last line of (\ref{cm1}))},
\end{eqnarray*}
yielding the second equation in (\ref{cm}).

\end{document}